\documentclass[runningheads]{llncs}
\usepackage[T1]{fontenc}
\usepackage{graphicx}
\usepackage{csquotes}
\usepackage{hyperref}

\usepackage{color}

\begin{document}
\title{Standardizing Access to Heterogeneous Quantum Backends: A Case Study on Cloud Service Integration with QDMI\vspace{-12pt}}
\titlerunning{Standardizing Access to Heterogeneous Quantum Backends}
\author{
    Patrick Hopf\inst{1,2} \and
    Sebastian Stern\inst{3} \and
    Robert Wille\inst{1,2} \and
    Lukas Burgholzer\inst{1,2} %
}
\authorrunning{P. Hopf et al.}
\author{
    Patrick Hopf\inst{1,2} \and
    Sebastian Stern\inst{3} \and
    Robert Wille\inst{1,2} \and
    Lukas Burgholzer\inst{1,2}\vspace*{-1mm}
}
\authorrunning{P. Hopf et al.}
\institute{
    Munich Quantum Software Company (MQSC), Garching near Munich, Germany
    \and
    Technical University of Munich, Munich, Germany
    \and
    Amazon Web Services, Inc., Seattle, WA 98109, USA\vspace*{2mm}
    \email{\{patrick,robert,lukas\}@munichquantum.software\\sterseba@amazon.com}\vspace*{-6mm}
}
\maketitle %
\begin{abstract} %
With an increasingly diverse portfolio of quantum backends, the adoption of standardized interfaces has become a key prerequisite for scalable access and interoperability within quantum software stacks.
The Quantum Device Management Interface (QDMI) addresses this challenge and is emerging as one of the de facto standards for hardware abstraction, enabling the unified management not only of individual Quantum Processing Units (QPUs) but also of complete full-stack cloud services.
This paper presents a case study demonstrating the integration of QDMI with Amazon Braket, a quantum computing cloud service that provides a single access point to a wide range of hardware technologies.
By treating the cloud service itself as a unified device, the proposed implementation enables management of the complete task lifecycle—ranging from authentication and circuit submission to result retrieval—across Braket’s heterogeneous set of simulators and hardware backends.
We detail the engineering insights gained from this integration and present a hands-on example workflow, ultimately paving the way for integrated access to cloud-hosted quantum resources from \emph{QDMI-enabled} software stacks.

\vspace*{-2mm}\vspace*{-3mm}
\end{abstract}
\section{Introduction}

Quantum computing is transitioning from isolated experimental systems toward tighter integration within production-grade High-Performance Computing (HPC) environments.
At the same time, the hardware ecosystem is rapidly diversifying, with superconducting, trapped-ion, neutral-atom, photonic, and simulator-based platforms exposed through heterogeneous vendor-specific interfaces.
For HPC centers and institutional software stacks, maintaining separate integration layers for multiple on-site devices and cloud providers is not scalable, particularly as hardware vendors and execution paradigms continue to evolve.
While quantum computing cloud services (such as Amazon Braket\footnote{Amazon Braket, \url{https://aws.amazon.com/braket/}}) simplify access to multiple technologies, their execution models, authentication mechanisms, and task semantics often differ from those of on-premise systems, complicating integration into existing quantum software stacks.

\begin{figure}[t]
    \centering
    \vspace*{-2mm}
    \includegraphics[width=0.95\textwidth]{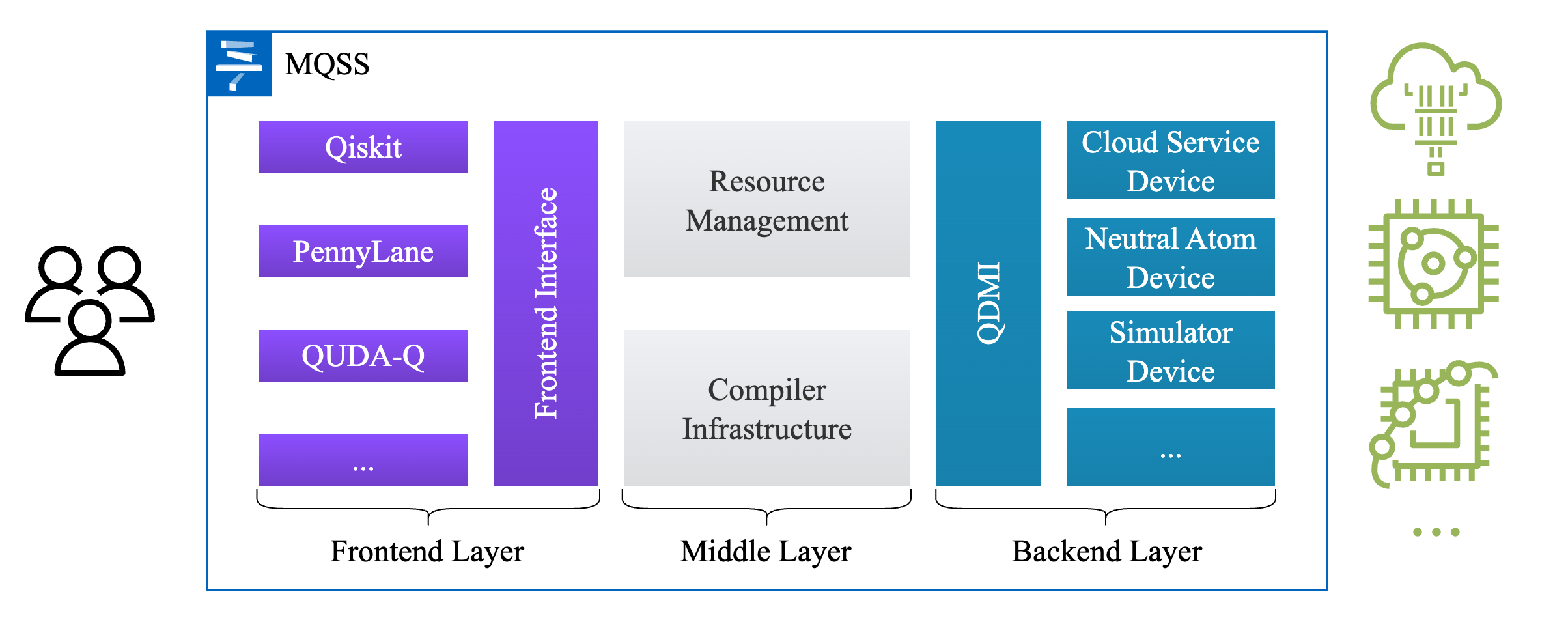}
    \caption{Illustration of the Munich Quantum Software Stack, showcasing the separation of concerns across frontend, middle, and backend layers. QDMI serves as a standardized abstraction for managing heterogeneous quantum backends, including local devices, simulators, and cloud-hosted resources.}
    \vspace*{-5mm}
    \label{fig:mqss}
\end{figure}

Stacks such as the Munich Quantum Software Stack (MQSS;~\cite{Burgholzer2026_MunichQuantumSoftwareStack}) or recent architectural proposals~\cite{SHEHATA2026107980} illustrate modular, hardware-agnostic approaches to integrating quantum resources into HPC environments.
These systems typically rely on layered designs that support portability, extensibility, and interoperability across heterogeneous backends~\cite{elsharkawyIntegrationQuantumAccelerators2025,humbleQuantumComputersHighPerformance2021}.
To assume this modularity, well-defined interfaces between user-facing programming environments, compilation and orchestration middleware, and hardware-specific execution layers are necessary.
The MQSS, for example, separates concerns across three main layers (see \autoref{fig:mqss}):
At the highest abstraction level, the \emph{frontend} layer provides user access via established programming frameworks and domain-specific libraries, expressing quantum programs in backend-independent forms.
The \emph{middle} layer handles compilation, optimization, scheduling, and resource management, transforming programs into hardware-compatible representations~\cite{willeMappingQuantumCircuits2019,willeMQTHandbookSummary2024}.
This step requires detailed knowledge of the target backend's capabilities---ranging from qubit connectivity to pulse-level control features~\cite{schoenbergerShuttlingScalableTrappedIon2024,stadeOptimalStatePreparation2025,Echavarria_Pulse}.
Finally, the \emph{backend} layer provides uniform access to heterogeneous QPUs and simulators.

To realize this separation, a well-defined boundary between the middle and backend layer is indispensable.
Without a unifying abstraction, each backend would require bespoke integration logic, thereby undermining portability and maintainability. %
Standardized backend abstractions are therefore becoming a prerequisite for scalable multi-platform environments.
The Quantum Device Management Interface (QDMI;~\cite{Wille_QDMI}) addresses this challenge by defining a uniform interface for device discovery, capability queries, job execution, and lifecycle management across heterogeneous quantum resources.

In principle, QDMI is general enough to represent not only individual QPUs and simulators but also cloud-hosted services that internally aggregate multiple heterogeneous devices.
However, this claim remains largely conceptual unless validated through concrete implementations.
It is unclear whether a multi-backend cloud service can be represented as a single logical device without introducing semantic mismatches or obscuring operational behavior.

Initial proposals such as Pilot-Quantum~\cite{ManthaPilotQuantum} provide unified workload management but operate primarily at the application level rather than exposing a unified backend representation within the middleware.
Similarly, deployments like Pawsey's Setonix-Q integrate on-premise and cloud resources but often rely on provider- or hardware-specific interfaces for execution.
Parallel community efforts such as the Quantum Resource Management Interface (QRMI;~\cite{sitdikov2025quantumresourcesresourcemanagement}) address related challenges by linking quantum devices to dedicated HPC schedulers, but focus primarily on resource management rather than device abstraction.

The central question motivating this work is therefore practical: can a real-world cloud service be integrated with QDMI while preserving both its abstraction guarantees and the operational constraints of the underlying platform?
To answer this, we present a case study on integrating Amazon Braket with QDMI, treating the entire cloud service as a single logical device.
Our contributions are:
\begin{itemize}
    \item a working open-source implementation exposing Amazon Braket backends through QDMI,
    \item a practical workflow demonstration covering the complete quantum job lifecycle from authentication to result retrieval,
    \item a systematic analysis of design decisions and abstraction (mis)matches encountered during the integration,
    \item and a generalizable blueprint that can guide analogous integrations for other cloud services.
\end{itemize}

By validating that cloud-native execution models can be incorporated into standardized backend abstractions, this work provides a concrete foundation for integrating heterogeneous cloud resources into QDMI-enabled quantum software stacks and HPC environments.
The remainder of this paper details the background on Amazon Braket and QDMI, describes the integration architecture, analyzes engineering insights, and provides an outlook on future developments.

\section{Background}

To keep this work self-contained, we briefly review the relevant background on Amazon Braket and QDMI before detailing the integration approach.

\subsection{Amazon Braket}

Amazon Braket is the quantum computing cloud service of Amazon Web Services (AWS) offering a single point of access to a variety of quantum computing technologies from multiple quantum hardware providers and a choice of quantum circuit simulators\footnote{Amazon Braket supported devices and regions, \url{https://docs.aws.amazon.com/braket/latest/developerguide/braket-devices.html}\label{footnote:BraketDevicesAndRegions}}.
Two access modes are available for QPUs: On-demand access requires no upfront commitment and requests to devices are queued and processed by the service in the order they were received.
With dedicated access, a QPU of choice can be reserved for exclusive use for a scheduled time period.
Braket provides features and abstractions enabling a consistent user experience across the different hardware modalities available.
The atomic request for processing a quantum program on a device is a \textit{quantum task}.
For gate-based devices, a program includes the quantum circuit description with measurement instructions and the number of shots.
On task creation, Braket typically compiles and optimizes submitted circuits to account for the qubit topology and native gate set of the target QPU.
However, for integration into a full software stack such as one based on QDMI, compilation is typically handled by the middle layer before the circuit reaches the backend.
To support such use cases, Braket offers \textit{verbatim} compilation, a feature that executes circuits exactly as defined without further modification.
Tasks are queued and scheduled for execution when the target device is available. Device availability, queue depth, and task queue position can be queried.
Braket is one of many different AWS cloud services and leverages some other services internally.
For instance, the results of computations performed on Braket are stored in a user-defined location in the object storage service Amazon S3\footnote{Amazon Simple Storage Service (S3), \url{https://aws.amazon.com/s3}}.

\begin{figure}[t]
    \centering
    \includegraphics[width=0.95\textwidth]{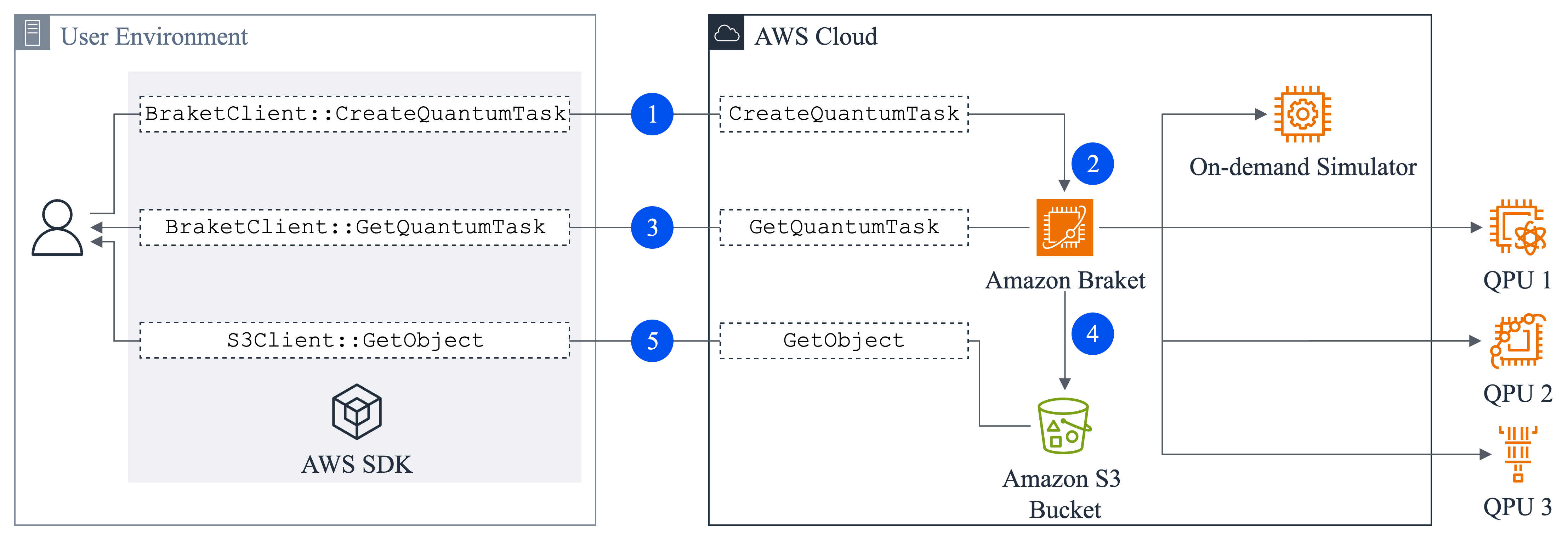}\vspace*{-3mm}
    \caption{High-level flow of running quantum programs on the Amazon Braket service from a local user environment through Braket and S3 client instances provided by an AWS SDK wrapping service APIs and handling the authentication of API calls: A quantum circuit is submitted for execution on a target device using \texttt{BraketClient::CreateQuantumTask} (1). The Braket service validates, compiles, optimizes, queues and schedules the circuit for execution on the target device (2). Task state and queue position can be queried with \texttt{BraketClient::GetQuantumTask} (3). Upon task completion, Braket stores the result in a user-defined location on Amazon S3 (4). The result object can be downloaded with \texttt{S3Client::GetObject} (5) using the object key returned in the \texttt{BraketClient::GetQuantumTask} response.}
    \vspace{-8pt}
    \label{fig:BraketTaskFlow}
\end{figure}

Braket offers a collection of APIs\footnote{Amazon Braket APIs, \url{https://docs.aws.amazon.com/braket/latest/APIReference}} for users to interact with the service via public endpoints accessible over the Internet.
An AWS account and security credentials for programmatic access are required to authenticate API requests.
To simplify service API calls from within a user application, AWS provides open-source SDKs for several programming languages including C++ and Python.
\autoref{fig:BraketTaskFlow} illustrates the end-to-end quantum task flow in Braket using an AWS SDK in a local environment to interact with the relevant service APIs.

While the Braket service natively accepts gate-based quantum programs in the OpenQASM 3 format---serving as a common intermediate representation---users typically develop applications in higher-level frameworks that offer more convenient abstractions.
Frameworks such as the Amazon Braket Python SDK\footnote{Amazon Braket Python SDK, \url{https://pypi.org/project/amazon-braket-sdk/}}, CUDA-Q\footnote{\href{https://nvidia.github.io/cuda-quantum/latest/using/backends/cloud/braket.html}{CUDA-Q} natively supports quantum kernel execution on Braket.}, PennyLane\footnote{PennyLane circuits can be executed on Braket with the \href{https://amazon-braket-pennylane-plugin-python.readthedocs.io/en/stable/}{PennyLane-Braket Plugin}.}, and Qiskit\footnote{Qiskit programs can be executed on Braket with the \href{https://qiskit-community.github.io/qiskit-braket-provider/}{Qiskit-Braket Provider}.} provide mechanisms to generate compatible OpenQASM 3 code for execution on Braket.

\subsection{QDMI}

QDMI is a C-based API that encapsulates device discovery, capability queries, job submission, monitoring, data retrieval, and lifecycle management for quantum systems in a backend-agnostic manner.
By design, the API is kept compact and extensible so that it can somewhat adapt to the rapidly evolving quantum hardware landscape without requiring (major) changes in higher layers of the software stack.
For instance, the API is designed to maintain binary compatibility across evolving implementations by hiding internal representations and implementation details behind opaque handles.

Any \emph{implementation} of the QDMI specification produces a so-called \emph{QDMI device} that encapsulates the management of one or more quantum backends.
The QDMI specification defines a layered lifecycle for device interactions.
The device itself follows this lifecycle pattern:
\begin{enumerate}
    \item \texttt{QDMI\_device\_initialize} initializes the device and prepares it for use.
    \item The initialized device is then used for subsequent interactions.
    \item \texttt{QDMI\_device\_finalize} finalizes the device and releases any associated resources when no longer needed.
\end{enumerate}

Once a device is initialized, the concept of device sessions is used to manage client interactions with the device and to encapsulate necessary context such as authentication state and target device information.
A \texttt{QDMI\_Device\_Session} generally has the following lifecycle:
\begin{enumerate}
    \item \texttt{QDMI\_device\_session\_alloc} creates a session (handle).
    \item \texttt{QDMI\_device\_session\_set\_parameter} configures session parameters such as authentication credentials and target device identifiers.
    \item \texttt{QDMI\_device\_session\_init} validates the session configuration and establishes a connection to the backend.
    \item The initialized session handle can then be used for subsequent device interactions such as capability queries and job management.
    \item \texttt{QDMI\_device\_session\_free} releases the session and associated resources when no longer needed.
\end{enumerate}
\pagebreak 

As pointed out above, sessions can be used to query device capabilities and properties at various levels of granularity.
This includes
\begin{enumerate}
    \item \texttt{QDMI\_device\_session\_query\_device\_property} for overall device-level capabilities and constraints,
    \item \texttt{QDMI\_device\_session\_query\_site\_property} for site- or qubit-level performance or quality metrics, and
    \item \texttt{QDMI\_device\_session\_query\_operation\_property} for operation- or gate-level performance or quality metrics.
\end{enumerate}
Such information is crucial for compilation and scheduling decisions in the middle layer of the software stack, enabling informed trade-offs between different backends without hard-coding specific details into higher layers.

The more important aspect for this work, however, is the job management interface\footnote{QDMI uses the notion of a \textit{job} for a quantum program execution corresponding to a \textit{task} on Amazon Braket. This should not be confused with a \textit{job} on Braket referring to the service-managed execution of a hybrid quantum-classical program with the \href{https://docs.aws.amazon.com/braket/latest/developerguide/braket-jobs.html}{Amazon Braket Hybrid Jobs} feature that is irrelevant for this work.}.
It allows clients to create and manage jobs on the device, which represent concrete program executions.
Jobs follow a similar lifecycle pattern:
\begin{enumerate}
    \item \texttt{QDMI\_device\_session\_create\_device\_job} creates a job (handle) from an initialized session.
    \item \texttt{QDMI\_device\_job\_set\_parameter} configures job parameters such as the quantum program and execution options.
    \item \texttt{QDMI\_device\_job\_submit} submits the job for execution on the device.
    \item \texttt{QDMI\_device\_job\_check} or \texttt{QDMI\_device\_job\_wait} can be used to track the job status asynchronously or synchronously, respectively.
    \item \texttt{QDMI\_device\_job\_get\_results} retrieves job results upon completion.
    \item \texttt{QDMI\_device\_job\_free} releases the job handle and any associated resources when no longer needed.
\end{enumerate}
Any implementation of QDMI must provide definitions for all aforementioned functions and adhere to the semantics defined in the specification.

\noindent Note that, due to the highly volatile nature of the quantum computing field, it can be expected that QDMI will have to evolve.
As such, the descriptions of the API and its semantics used in this work are based on the current version of the QDMI specification (version 1.2 as of this writing), which is subject to change as the standard matures and adapts to new requirements and use cases. %
However, the core concepts and lifecycle patterns described in this work are reasonably expected to remain relevant and applicable even as the API evolves, given that they are designed to be general and adaptable to a wide range of quantum backends and execution models.
For an up-to-date and comprehensive reference on the API and its semantics, please refer to the live QDMI specification\footnote{QDMI specification, \url{https://github.com/Munich-Quantum-Software-Stack/QDMI}.}.

\section{Integration}

Having established the context of Amazon Braket and the QDMI specification, we now turn to the core contribution of this work: the practical integration of the cloud service into the standardized interface.
The primary challenge lies in bridging the conceptual gap between a stateful, session-oriented hardware abstraction (QDMI) and a stateless, request-based cloud API (AWS).
To resolve this, we model the entire Amazon Braket service as a single logical \emph{QDMI device}.
This design choice allows any client using the QDMI device to maintain local state where necessary---such as authentication contexts and job tracking---while delegating actual execution to the cloud infrastructure.
The following sections detail the resulting architecture, focusing on how the strict lifecycle of QDMI maps onto the flexible, on-demand nature of cloud resources.

\begin{figure}
    \centering
    \includegraphics[width=\textwidth]{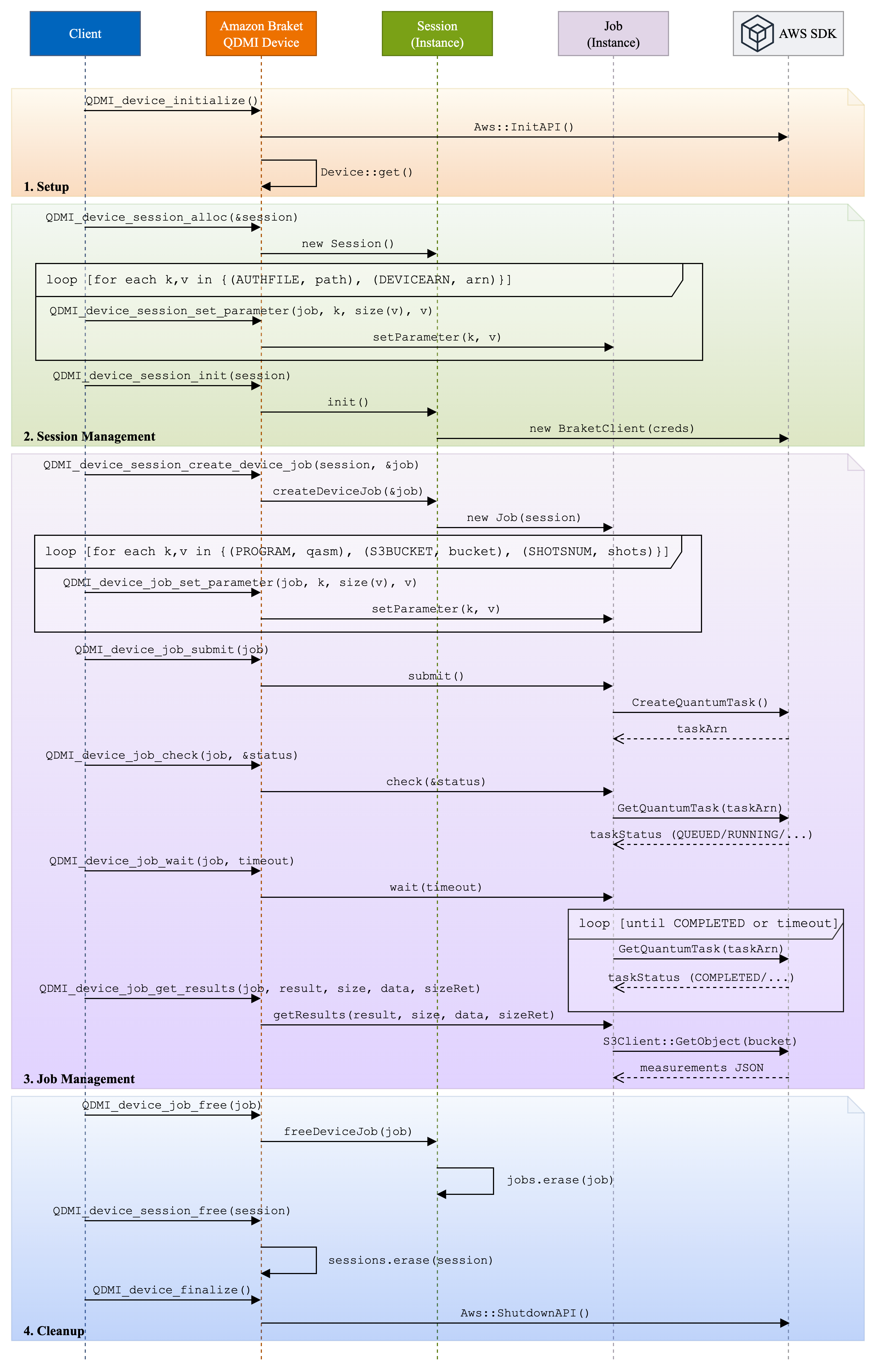}
    \caption{Sequence diagram illustrating the interaction between the client and the Amazon Braket QDMI Device across the stages of setup, session management, job management, and cleanup. Some intuitive responses are omitted for readability.}
    \label{fig:sequence-diagram}
\end{figure}

\subsection{The General Workflow}
To understand the runtime behavior of the integration, it is instructive to follow the complete lifecycle of a quantum task.
As visualized in \autoref{fig:sequence-diagram}, the interaction between a QDMI client and our implementation follows a structured sequence that spans device initialization, session management, job execution, and final teardown.

\subsubsection{1. Setup:}
The Device is initialized via \texttt{QDMI\_device\_initialize}, which ensures that the AWS SDK is initialized exactly once via \texttt{Aws::InitAPI} in a thread-safe manner.
Afterwards, a process-wide \texttt{Device} singleton is constructed, which manages session lifecycles, maintains a cache of device metadata to avoid redundant API calls, and provides shared infrastructure such as job identifier generation.
The setup phase therefore establishes a consistent runtime environment and the coordination layer required for all subsequent session and job operations.

\subsubsection{2. Session Management:}
Each client interaction begins by allocating an isolated device session via \texttt{QDMI\_device\_session\_alloc}, which creates a session instance tracked by the global device singleton.
The session is then configured through \texttt{QDMI\_device\_session\_set\_parameter}, including the Amazon Resource Name (ARN) for the target device (see \textit{Device ARN} in \autoref{footnote:BraketDevicesAndRegions}) and AWS security credentials\footnote{AWS security credentials, \url{https://docs.aws.amazon.com/IAM/latest/UserGuide/security-creds.html}\label{footnote:SecurityCredentials}} for request authentication and authorization provided either through a credentials file or via explicit access-key parameters.
Finally, after validating the configuration and resolving defaults where necessary, \texttt{QDMI\_device\_session\_init} constructs a session-specific instance of \texttt{Aws::Braket::BraketClient}, thereby encapsulating client identity without affecting global device state.
\pagebreak

\subsubsection{3. Job Management:}
A QDMI job is created from the initialized session via \texttt{QDMI\_device\_session\_create\_device\_job}. 
It must be configured with a (gate-based) quantum program in OpenQASM 3 format, a custom output Amazon S3 bucket, and a number of shots using \texttt{QDMI\_device\_job\_set\_parameter}.
For users with dedicated access, a Braket reservation ARN can be provided optionally.
Submitting the configured job with \texttt{QDMI\_device\_job\_submit} internally calls
\texttt{BraketClient::CreateQuantumTask}, which returns a unique task identifier.
\texttt{QDMI\_device\_job\_check} or \texttt{QDMI\_device\_job\_wait} subsequently use this identifier to query the task status or wait for the job to complete by invoking \texttt{BraketClient::GetQuantumTask}.
After completion, the measurement results are retrieved via \texttt{QDMI\_device\_job\_get\_results} by fetching the corresponding JSON object from the previously set S3 bucket through \texttt{S3Client::GetObject}.

\subsubsection{4. Cleanup:}
Resources are released in a teardown sequence by first deallocating the job with \texttt{QDMI\_device\_job\_free}, then freeing the session with \texttt{QDMI\_device\_session\_free}, and finally calling \texttt{QDMI\_device\_finalize}, which invokes \texttt{Aws::ShutdownAPI} to cleanly release AWS SDK resources and ensure that no further device calls can occur.

\subsection{Implementation and Distribution}
The integration is realized as an open-source software project written in modern C++ (C++20) to leverage recent language features for type safety and performance\footnote{Implementation freely available at \url{https://github.com/munich-quantum-software/amazon-braket-qdmi-device}.}. %
The architecture revolves around the QDMI library, which is utilized to generate namespace-isolated headers for the specific device implementation, ensuring symbol safety when multiple QDMI devices are loaded simultaneously.
The core logic implements the required QDMI interface functions by mapping them to the AWS SDK for C++, which serves as the primary dependency for communicating with the Braket service.

To ensure broad portability and ease of deployment in diverse HPC environments, the device is compiled as a shared library that statically links the AWS C++ SDK.
This design choice results in a self-contained binary module that minimizes runtime dependency conflicts—a crucial consideration for integration into complex software stacks.

Facilitating adoption, the implementation is distributed as a binary Python wheel, allowing for installation via standard package managers (e.g., \texttt{uv pip install amazon-braket-qdmi}).
This package includes a command-line interface that exposes the location of the library, include headers, and CMake configuration files, enabling downstream build systems to locate and link against the device implementation without manual configuration.
Finally, the project employs a rigorous development workflow, utilizing GitHub Actions for continuous integration (CI) and automated linting pipelines to maintain high code quality standards.

\section{Integration Insights}

Overall, the integration of Amazon Braket into the QDMI framework proved to be remarkably seamless. %
The conceptual alignment between the task-based execution model of the cloud service and the job-based lifecycle of QDMI allowed for a direct mapping of most core functionalities.
In particular, the asynchronous nature of cloud task submission naturally fits the non-blocking job submission and polling mechanisms defined by QDMI.
However, adapting the stateless cloud API to the stateful hardware interface required specific handling, particularly for status reporting and AWS regions, while some advanced features of Braket remain outside the scope of the current QDMI specification.

\subsubsection{Mapping Status Models}
\begin{table}[t]
    \centering
    \caption{Mapping between QDMI job states and Amazon Braket task states (top), and between their respective device states (bottom).}
    \vspace{-4pt}
    \begin{tabular}{l|l}
        \textbf{QDMI} & \textbf{Amazon Braket} \\
        \hline
        \texttt{CREATED}  & \texttt{CREATED} \\
        \texttt{QUEUED}   & \texttt{QUEUED} \\
        \texttt{RUNNING}  & \texttt{RUNNING} \\
        \texttt{DONE}     & \texttt{COMPLETED} \\
        \texttt{FAILED}   & \texttt{FAILED} \\
        \texttt{CANCELED} & \texttt{CANCELLED} \\
        no corresponding concept \, & \texttt{CANCELLING} \\
        no corresponding concept  & \texttt{NOT\_SET} \\
        \hline
        \texttt{ERROR}        & no corresponding concept \\
        \texttt{OFFLINE}      & \texttt{RETIRED} \\
        \texttt{IDLE}         & \texttt{ONLINE} $+$ (Queue $<$ threshold) \\
        \texttt{BUSY}         & \texttt{ONLINE} $+$ (Queue $\geq$ threshold)\\
        \texttt{MAINTENANCE}  & \texttt{OFFLINE} \\
        \texttt{CALIBRATION}  & \texttt{OFFLINE} \\
        no corresponding concept  & \texttt{NOT\_SET} \\
        \hline
    \end{tabular}
    \label{tab:status-mapping}
    \vspace{-8pt}
\end{table}

The most significant semantic mismatch appears in the job and device status definitions (see \autoref{tab:status-mapping}).
QDMI uses a compact set of six states for jobs, whereas Amazon Braket exposes eight, including transitional states like \texttt{CANCELLING} and \texttt{NOT\_SET}.
To bridge this, we treat \texttt{CANCELLING} as a running state until terminal resolution, and raise an \texttt{ERROR} when encountering \texttt{NOT\_SET}.
Conversely, QDMI's device status model is more fine-grained than Braket's regarding operational availability (distinguishing \texttt{IDLE} vs. \texttt{BUSY} and \texttt{MAINTENANCE} vs. \texttt{CALIBRATION}).
Since Braket only exposes \texttt{ONLINE}, \texttt{OFFLINE}, and \texttt{RETIRED}, we map \texttt{OFFLINE} to \texttt{MAINTENANCE} and implement a queue-depth threshold to heuristically differentiate between \texttt{IDLE} and \texttt{BUSY} for online devices.
This logic, while necessary, highlights a fundamental difference between the explicit resource control of on-premise systems and the abstracted multi-tenant nature of cloud services.

\subsubsection{Service Granularity and Regionality.}
AWS services are hosted in separate geographic locations (Regions) and Amazon Braket does not automatically transfer data between them.
However, QDMI abstracts away physical location, assuming a unified device handle.
To reconcile this, our implementation dynamically determines the requisite AWS Region from the device ARN during session initialization.
It then instantiates the internal AWS SDK client specifically for that region.
Since Amazon Braket requires an explicit storage location for task results within the same region, the implementation also leverages QDMI's extensible parameter interface.
Even though QDMI does not natively know about the concept of AWS Regions or S3 buckets, it allows for custom parameters to be defined and passed through the session and job configuration.
For instance, the device implementation uses a \texttt{CUSTOM} job parameter for the S3 bucket name, which users can set when configuring their job.

\subsubsection{Advanced Execution Capabilities.}
Finally, it is worth noting that Amazon Braket offers advanced execution features that extend beyond the capabilities of the current QDMI specification (v1.2).
These include \textit{\href{https://docs.aws.amazon.com/braket/latest/developerguide/braket-constructing-circuit.html\#braket-program-set}{program sets}}, which allow multiple quantum circuit executions to be processed in a single quantum task to minimize overhead, and \textit{\href{https://docs.aws.amazon.com/braket/latest/developerguide/braket-pulse-control.html}{pulse control}}, which provides pulse-level access to supported devices enabling execution of quantum programs specified with the analog instructions that control the qubits of a QPU.
While these features are not yet exposed through QDMI, there is no fundamental architectural barrier to their inclusion.
Indeed, community discussions on extending QDMI with pulse-level support are already underway~\cite{Echavarria_Pulse}, suggesting that future iterations of the standard can naturally incorporate these capabilities.

\section{Outlook and Conclusion}

In conclusion, this case study successfully demonstrates the integration of the Amazon Braket cloud service into the standardized Quantum Device Management Interface (QDMI).
This confirms that QDMI is general enough to support not only local accelerators but also multi-regional cloud backends within a single interface.

With this integration in place, future research can shift from basic connectivity to system-level optimization.
A key direction is investigating the latency and data-movement overheads involved in hybrid workflows that span on-premise HPC resources and cloud QPUs.
Furthermore, the availability of a cloud-backed QDMI device enables the development of unified schedulers that can dynamically broker workloads between local accelerators and cloud backends based on availability, queue depth, or budget constraints.
By open-sourcing this implementation, we aim to lower the barrier for such experiments and accelerate the adoption of truly hybrid, multi-provider quantum computing environments.

\newpage
\bibliographystyle{splncs04}
\bibliography{header,refs}
\end{document}